\documentclass[a4paper,eqsecnum,12pt,nofootinbib]{revtex4}

\usepackage{amssymb}
\usepackage{amsfonts}
\usepackage{amsmath}
\usepackage{mathrsfs} % for \mathscr{}
\usepackage{color}
\usepackage{fancyhdr}
\usepackage{enumerate}

\newcommand{\omits}[1]{}

\definecolor{dyellow}{rgb}{1.,0.8,.0}
\definecolor{myblue}{rgb}{.1,.1,.7}
\definecolor{dcyan}{rgb}{.0,.6,.6}
\definecolor{dmagenta}{rgb}{0.6,0.0,0.6}
\definecolor{brown}{rgb}{0.6,0.2,0.}
\definecolor{darkblue}{rgb}{.0,.0,0.5}
\definecolor{darkred}{rgb}{0.75,0.0,0.0}
\definecolor{orange}{rgb}{1.,.6,.0}
\definecolor{dorange}{rgb}{0.8,.4,.0}
\definecolor{green}{rgb}{0.0,1.0,0.0}
\definecolor{darkgreen}{rgb}{0.0,0.7,0.0}
\definecolor{lightgrey}{rgb}{0.7,0.7,0.7}
\definecolor{purple}{rgb}{.4,.0,.4}
\begin{document}
\hyphenpenalty=1000

\title{Reformulation of Boundary BF Theory Approach to Statistical Explanation of the Entropy of Isolated Horizons}

\author{Chao-Guang Huang}
\email{huangcg@ihep.ac.cn}
\author{Jingbo Wang}
\email{ wangjb@ihep.ac.cn}
\bigskip

\affiliation{Institute of High Energy Physics and Theoretical Physics Center for
Science Facilities, \\ Chinese Academy of Sciences, Beijing, 100049, People's Republic of China}

\begin{abstract}
It is shown in this paper that the symplectic form for the system consisting of $D$-dimensional bulk Palatini gravity and SO$(1,1)$ BF theory on an isolated horizon as a boundary just contains the bulk term.   An alternative quantization procedure for the boundary BF theory is presented. The area entropy is determined by the degree of freedom of the bulk spin network states which satisfy a suitable boundary condition.  The gauge-fixing condition in
the approach and the advantages of the approach are also discussed.
\end{abstract}

%\pacs{04,50,-h}
%\keywords{ Loop quantum gravity, higher dimensional rotating isolated horizon, BF theory} \preprint{} \dedicated{} %\maketitle

\maketitle
\tableofcontents

%\flushbottom

%%%%%%%%%%%%%%%%%%%%%%%%%%%%%%%%%%%%%%%%%%%
%%%%%%%%%%%%%%%%%%%%%%%%%%%%%%%%%%%%%%%%%%%
%%%                                     %%%
%%%          Introduction               %%%
%%%                                     %%%
%%%%%%%%%%%%%%%%%%%%%%%%%%%%%%%%%%%%%%%%%%%
%%%%%%%%%%%%%%%%%%%%%%%%%%%%%%%%%%%%%%%%%%%

\section{Introduction}
It is well known that black holes behave as thermodynamic systems, having temperature \cite{hawk1} and entropy \cite{bk1}. Then a natural question arises: what are the underlying microscopic degrees of freedom. The first breakthrough is the work of Strominger and Vafa on extremal black holes in string theory \cite{sv1}. Later on their results extend to a wide variety of extremal and near-extremal black holes \cite{dm1}. Another explanation is from loop quantum gravity, in which inequivalent spin networks crossing the horizon account for the entropy \cite{kras1,rov2}. The more careful variation of this idea \cite{abck,abk1} use the Chern-Simons theory on the punctured manifold to describe the microscopic degrees of freedom on the boundary. There are other theories to explain the entropy of the black holes, such as entanglement entropy \cite{ent1,ent2}, ``heavy" states in induced gravity \cite{ind1}, Carlip's 2D conformal field theory approach \cite{carlip2} and so on. For a brief review see \cite{carlip0}.

Isolated horizons (IHs) \cite{abk1,afk} are the generalization of event horizons
of black holes.  They have many applications in mathematical physics, numerical
relativity and quantum gravity \cite{abf}.  They provide more physical setting for
studying the statistical origin of the entropy of a black hole
in loop quantum gravity \cite{{abk2},{enpp1}}. The calculation of the entropy is based on counting the dimension of
Hilbert spaces of the boundary Chern-Simons theory on the IH \cite{abck,rot1,rot2}. In literature, there is another way to calculate the entropy of the black hole in 4 dimension, beginning with Rovelli's work \cite{rov2}. It attributes the entropy to the degrees of freedom in the bulk spin network states related to its area, and doesn't use the boundary field theory.

Recently, a new approach to calculate the entropy of IHs in the framework of loop quantum gravity is proposed \cite{{wmz},{wh1}}.  In this approach, the entropy of an IH is attributed
to the number of degrees of freedom of the quantized SO(1,1) BF theory instead of the Chern-Simons theory. The procedure is the same as the Chern-Simons theory approach: starting form the Palatini action.  One first analyses the symplectic form to get the boundary BF theory, and then quantizes the bulk and boundary BF theory separately to get the full Hilbert space as the tensor product of bulk and boundary Hilbert space, and finally applies the quantized boundary condition to get the permissible boundary states. The number of the independent permissible boundary states which satisfy some constraints will give the entropy of an IH.

In this paper, we reformulate the boundary BF theory approach. A boundary BF action is added to the original Palatini action, which cancel the boundary symplectic current. The obtained symplectic form has only the bulk term, so the final Hilbert space is also just the bulk Hilbert space which spanned by the spin network states from loop quantum gravity.  In the paper,
a new way to quantize the boundary BF theory will also be presented.
The key observation is that the boundary BF theory is actually a pure BF theory coupled
to a bulk Palatini gravity.  Classically, the $B$ field in the boundary BF theory
can be decomposed into a closed form just like the $B$ field in a pure
BF theory and a fixed, non-closed form determined by the coupling.  The former
is a topological field theory on the IH without local degree of freedom, the latter defines
the local degree of freedom on the IH via the coupling with the bulk Palatini gravity.
After quantization,
the pure BF theory sets up the complete basis of the boundary Hilbert space \cite{bf1} and the boundary BF theory provides the boundary condition to choose the bulk spin network states. The number of the independent spin network states which satisfying
the condition will give the entropy of the IH.

The paper is organized as follows.  In section 2, the asymptotical behavior of the symplectic form for the $D$-dimensional Palatini gravity near an IH is analyzed. In section 3, the approach of quantization of a BF theory coupled to bulk
Palatini gravity on an IH
is presented. In section 4, the bulk quantum states are discussed and the entropy is given. In section 5,  the relation between the conclusion and the choice of the residual gauge symmetry of IHs is studied.
The concluding remarks will be made
in section 6.
Throughout the paper, we use the units of $\hbar=c=1$.

%%%%%%%%%%%%%%%%%%%%%%%%%%%%%%%%%%%%%%%%%%%
%%%%%%%%%%%%%%%%%%%%%%%%%%%%%%%%%%%%%%%%%%%
%%%                                     %%%
%%%        Isolated Horizons            %%%
%%%                                     %%%
%%%%%%%%%%%%%%%%%%%%%%%%%%%%%%%%%%%%%%%%%%%
%%%%%%%%%%%%%%%%%%%%%%%%%%%%%%%%%%%%%%%%%%%

{\section{Isolated horizon as an internal boundary}

Consider a $D$-dimensional asymptotically-flat spacetime ${\cal M}$ bounded by an
IH, denoted by $\Delta$, and suppose that it
be described by the Palatini action for the interior of ${\cal M}$ and
an SO$(1,1)$ BF theory for $\Delta$.  That is,
 \begin{equation}\label{1}
  S=-\frac{1}{2\kappa}\int_{\mathcal{M}} \Sigma_{IJ}   \wedge F^{IJ} +\int_{\Delta}{\cal B}\wedge {\rm d} {\cal A},
\end{equation}
where $\kappa\equiv 8\pi G$,
\begin{equation}\label{Sigma}
\Sigma_{IJ}=\dfrac 1 {(D-2)!}\varepsilon_{IJK\cdots L} e^K\wedge \cdots \wedge e^L,
\end{equation}
$e^I$ is the orthogonal co-vielbein, $\varepsilon_{IJK\cdots L}$ is the Levi-Civita symbol, $F^{IJ}={\rm d}A^{IJ}+A^I{}_K\wedge A^{KJ}$
is the curvature of the SO$(D-1,1)$ connection 1-form $A^{IJ}$ on ${\cal M}$, ${\cal A}$ is the SO$(1,1)$ connection on $\Delta$, and ${\cal B}$ is the $B$-field in BF theory. The spacetime region ${\cal M}$ is supposed to be bounded by the initial and final spacelike
hypersurfaces $M_1$ and $M_2$ and an isolated horizon $\Delta$ from the inner, and to extend to
spatial infinity $i^0$.  The cross section of $\Delta$ with the hypersurface $M_1$ and $M_2$ are denoted by $K_1$ and $K_2$. All fields are assumed to be smooth and satisfy the standard
asymptotic boundary condition at spatial infinity, $i^0$.

Let $l^a$ be a null vector and normal to $\Delta$ on $\Delta$,
$e_{\tt A}^a\ ({\tt A}=2, \cdots, D-1)$ be $D-2$ spacelike vectors orthogonal
to $l^a$ and tangent to a section of $\Delta$ on $\Delta$, and $n^a$ be the future-directed null vector field such that
$\langle l, n\rangle \triangleq -1$.  Let $(v, \zeta^i,\ i=2, \cdots, D-1)$
be coordinates on $\Delta$ such that ${\cal L}_l v \triangleq 1$.
Define $r$ coordinate via ${\cal L}_n r = -1$ and $r=0$ on $\Delta$.
Lie drag of $(v, \zeta^i)$ along $n^a$ establishes a system of coordinates $(v, r, \zeta^i)$
near the IH, called the Bondi-like system of coordinates \cite{abdf}. The NP null co-vielbein $(l_a, n_a, e^{\tt A}_a)$ can be expressed in terms of
the Bondi-like coordinates.
Those null co-vielbein defines
the following orthogonal co-vielbein:
\begin{equation}\label{6}
\begin{aligned}
&   e^0=\sqrt{\frac{1}{2}}(\alpha n+\frac{1}{\alpha} l),\ &&
    e^1=\sqrt{\frac{1}{2}}(\alpha n-\frac{1}{\alpha} l),\quad
    &     e^{\tt A},
\end{aligned}
\end{equation}
where $\alpha(x)$ is an arbitrary function of the coordinates. Restricted to the IH $\Delta$, $l\triangleq 0$ so that
\begin{equation}\label{7}\begin{split}
    e^0\triangleq e^1.
\end{split}\end{equation}
After some straightforward calculation, the following conditions are obtained:
\begin{equation}\label{7a}\begin{split}
\Sigma_{0{\tt A}}\triangleq -\Sigma_{1{\tt A}},\,A^{0{\tt A}}\triangleq A^{1{\tt A}},\\
  A^{01}\triangleq (\tilde \kappa {\rm d} v+{\rm d}\ln \alpha)+(\delta^{\tt A B} \pi_{\tt A} e_{\tt B})  \triangleq : \bar A^{01}+ \tilde A^{01}.
\end{split}\end{equation}

The variation of the action (\ref{1}) will give rise to the vacuum Einstein field equations in the bulk and
\begin{equation}\label{2}
    {\rm d} {\cal{B}} \triangleq \frac{\Sigma_{01}}{\kappa},\qquad {\rm d}{\cal{A}}\triangleq 0
\end{equation}
on the boundary if one identities the SO$(1,1)$ connection ${\cal{A}}$ in BF theory with the nonrotating part $\bar{A}^{01}$ of $A^{01}$ on the IHs. Also, one can get the symplectic potential density which contain two terms,
\begin{equation}\label{3}
    \theta({\delta})=\theta_M({\delta})+\theta_K({\delta})=(-1)^{D-1}(\frac{1}{2\kappa}\Sigma_{IJ}\wedge {\delta} A^{IJ}+{\cal{B}}\wedge \delta {\cal{A}}).
\end{equation}
The second-order exterior variation will give the symplectic current,
\begin{equation}\label{4}
    J({\delta}_1,{\delta}_2)=(-1)^{D-1}(\frac{1}{\kappa}{\delta}_{[2}\Sigma_{IJ}
    \wedge {\delta}_{1]} A^{IJ}+{\delta}_{[2}{\cal{B}}
    \wedge {\delta}_{1]} {\cal{A}}).
\end{equation}
The nilpotent of exterior variation, $\delta^2=0$, implies $d J=0$.
Applying Stokes' theorem to the integration
$\int_{\mathcal{M}} dJ=0$, one can get the following equation:
\begin{eqnarray}\label{5}\begin{split}
   \frac{1}{\kappa}(\int_{M_2}\delta_{[2}\Sigma_{IJ}\wedge \delta_{1]} A^{IJ}
   -\int_{M_1}\delta_{[2}\Sigma_{IJ}\wedge \delta_{1]} A^{IJ}%\\
   -\int_{\Delta}\delta_{[2}\Sigma_{IJ}\wedge \delta_{1]} A^{IJ})\\
   +\int_{K_2}\delta_{[2}{\cal{B}}\wedge \delta_{1]}{\cal{A}}
   -\int_{K_1}\delta_{[2}{\cal{B}}\wedge \delta_{1]} {\cal{A}} =0 .
\end{split}
\end{eqnarray}

It has been show firstly for nonrotating IHs \cite{wh1} and then for rotating IHs \cite{wh2} that
 \begin{equation}\label{8}
\frac{1}{\kappa}\int_{\Delta} \delta_{[2}\Sigma_{IJ}\wedge\delta_{1]} A^{IJ}=
\int_{K_2} \delta_{[2}{\cal{B}} \wedge \delta_{1]}{\cal{A}}-
\int_{K_1} \delta_{[2}{\cal{B}}\wedge \delta_{1]}{\cal{A}}.
\end{equation}
So
 \begin{equation}\label{9}
   \Omega(\delta_1,\delta_2)=\frac{1}{\kappa}\int_{M}\delta_{[2}\Sigma_{IJ}\wedge \delta_{1]} A^{IJ}
\end{equation}
is independent of $M$ and can be considered as the symplectic form.  Notice that since the boundary symplectic current cancel with each other, the $M$ can be considered as an open region without an internal boundary.

 \bigskip

%%%%%%%%%%%%%%%%%%%%%%%%%%%%%%%%%%%%%%%%%%%
%%%%%%%%%%%%%%%%%%%%%%%%%%%%%%%%%%%%%%%%%%%
%%%                                     %%%
%%%            BF theory                %%%
%%%                                     %%%
%%%%%%%%%%%%%%%%%%%%%%%%%%%%%%%%%%%%%%%%%%%
%%%%%%%%%%%%%%%%%%%%%%%%%%%%%%%%%%%%%%%%%%%

\section{SO$(1,1)$ boundary BF theory}
\setcounter{equation}{0}
In $(D-1)$-dimensional spacetime $\Delta$, the action of an ordinary SO$(1,1)$ BF theory
can be written as \cite{bf2,bf1}
\begin{equation}\label{22}
    S[{\cal B},{\cal A}]=\int_{\Delta}\textrm{Tr}( {\cal B}\wedge {\cal F}({\cal A}))=\int_{\Delta}{\cal B}\wedge {\rm d}{\cal A}.
\end{equation}
where ${\cal A}$ is an SO$(1,1)$ connection field, ${\cal F}$ its field strength 2-form, and
${\cal B}$ a $(D-3)$-form field in the adjoint representation of SO$(1,1)$.
From the action (\ref{22}), one can easily obtain the field equations as
\begin{equation}\label{22a}
    {\cal F}:={\rm d}{\cal A}=0,\qquad {\rm d}{\cal B}=0.
\end{equation}
In the BF theory, $A$ is a flat connection, $B$-field is closed.

On the other hand, the field equations for the boundary BF theory we need on the IH
are
\begin{equation}\label{27}
    F={\rm d} {\cal A}=0,\qquad {\rm d}{\cal B} \triangleq\frac{\Sigma_{01}}{\kappa}.
\end{equation}
Compared with (\ref{22a}), equation (\ref{27}) shows that ${\cal A}$ remains a flat connection
while ${\cal B}$-field is no longer closed
and that the bulk field $\Sigma_{01}$ serves as the source of the ${\cal B}$ field, locally.
The canonical momentum conjugate to ${\cal A}$ is still
\begin{equation}
{\cal B} =\frac{\partial {\cal L}}{\partial \dot{\cal A}},
\end{equation}
which plays the role of the `electric' field for the SO(1,1) gauge potential ${\cal A}$.
Then, the second equation of (\ref{27}) is interpreted as the Gauss's law with the external
source of
${\Sigma_{01}}/{\kappa}$, in analogous to the equation in electromagnetism with nonvanishing charge density.
The symplectic structure on the phase space is given by
\begin{equation}
\Omega({\cal A}, {\cal B};\delta_1,\delta_2)=\int_S \delta_{[1}{\cal A}\wedge \delta_{2]}{\cal B},
\end{equation}
where $S$ is the $(D-2)$-dimensional spacelike hypersurface.
The second field equation of (\ref{27}), as the Gauss's constraint, generates the action of gauge
 transformations on the phase space of initial data ${\cal A}$ and ${\cal B}$.  The first equation
 of (\ref{27}) is analogous to an equation requiring the `magnetic' field to vanish.   As a constraint, the first field equation generates the gauge transformation,
\begin{equation}\label{newgaugetransf}
{\cal A} \mapsto {\cal A}, \quad {\cal B} \mapsto {\cal B}+{\rm d}\phi,
\end{equation}
on the phase space.  In other words, the action of  a boundary BF theory (coupled to GR), like a pure BF theory, is invariant under the above two kinds of gauge transformations.
For simplicity, we may split ${\cal B}$ into two parts.
One is denoted by ${\cal B}_{\rm es}$, which is determined by the external source.  Remember that
$\Sigma_{01}$ is a $v$-independent $(D-2)$-form on an IH because
$l^a \nabla_a \Sigma_{01} \triangleq \frac d {dv} \Sigma_{01} = 0$.
Without loss of generality, ${\cal B}_{\rm es}$ may be assumed to be a $v$-independent $(D-3)$-form
on an IH and fixed under the above gauge transformation.  The other part is
${\cal B}_{\rm c}$ which is always closed, i.e.,
\begin{equation}
{\rm d}{\cal B}_{\rm c} =0.
\end{equation}
The gauge transformation (\ref{newgaugetransf}) becomes
\begin{equation}\label{newgaugetrans2}
{\cal A} \mapsto {\cal A}, \qquad {\cal B}_{\rm es} \mapsto {\cal B}_{\rm es}, \qquad {\cal B}_{\rm c} \mapsto {\cal B}_{\rm c} +{\rm d}\phi ,
\end{equation}
and the Lagrangian is rewritten as
\begin{equation}
\mathscr{L}_{\rm BF}={\cal B}\wedge {\cal F} ={\cal B}_{\rm c}\wedge {\cal F} +{\cal B}_{\rm es}\wedge {\cal F}.
\end{equation}
The former term looks like a pure BF theory, playing the role of a `free' BF theory, which has
no local degree of freedom.
The latter term is the interaction term between BF theory and gravity, which provides the local
degree of freedom (for the boundary).

Since the gauge group $G=$SO$(1,1)$ is an Abelian, it is easy to construct a gauge-invariant
function of ${\cal B}_{\rm c}$, as an observable on the gauge-invariant Hilbert space, by the integral
\begin{eqnarray}
  \int_{K_i} {\cal B}_{\rm c},
\end{eqnarray}
where $K_{i}$ is a closed, oriented $(D-3)$-dimensional submanifold in $S$.
It is the flux of the `electric' field ${\cal B}_{\rm c}$ through $K_i$.  Recall that
the spatial section of an IH is closed and thus has no boundary.
We have
\begin{equation}\label{Bc-BoundCond}
0=\int_S {\rm d}{\cal B}_{\rm c} = \sum_{P_i} \int _{P_i}{\rm d}{\cal B}_{\rm c}
-\sum_{P_i \bigcap P_j} \int_{P_i \bigcap P_j} {\rm d}{\cal B}_{\rm c}
\end{equation}
if there is no intersection of 3 or more patches.  After triangulation of $S$, it becomes
\begin{equation}\label{Bc-BoundCond-Triang}
0=\int_S {\rm d}{\cal B}_{\rm c} = \sum_{\alpha} \int _{s_\alpha} {\rm d}{\cal B}_{\rm c} =\sum_{\alpha} \oint_{\eta_{\alpha}}{\cal B}_{\rm c},
\end{equation}
where $s_\alpha$ is a $(D-2)$-simplex, $\eta_{\alpha}=\{\eta_{\alpha 1}, \cdots \eta_{\alpha (D-1)}\}$ is a set of $(D-1)$ pieces of oriented-compatible $(D-3)$-simplices surrounding $s_\alpha$, the summation is over all $(D-2)$-simplices.   The triangulation $\{s_\alpha\}$
for $S$ may be regarded as a partition of $S$ because each simplex $s_\alpha$ is a closed set with open interior.
(The intersection of $s_\alpha$ and $s_\beta$ in a triangulation is, by definition, just
a simplex of 1 dimension lower than $s_\alpha$.)  Since ${\cal B}_{\rm c}$ is closed and each simplex
is topological trivial, each term $\oint_{\eta_{\alpha}}{\cal B}_{\rm c}$
is equal to zero.  It is remarked that for a given triangulation the `area' of each simplex $s_\alpha$ must be greater than 0 with respect to the flat metric on the simplex.

The loop quantization of a pure BF theory has been presented in Ref. \cite{bf1}.
The basis of the Hilbert space is provided by a set of generalized `spin network states'.
Each `spin network state' ${\cal S}$ of the pure BF theory in a $(D-1)$-dimensional spacetime
may be used to define a triangulation ${\cal T}$ for a $(D-2)$-dimensional hypersurface $S$
in the following way.  Each
$(D-2)$-simplex $s_\alpha$ of the triangulation ${\cal T}$ contains one and only one vertex $V_\alpha$
of the `spin network state' ${\cal S}$.  Then, the summation in (\ref{Bc-BoundCond-Triang}) will be different for different `spin network states'.

Now, choose $S=H$, a section of an IH.  After triangulation, the expression
\begin{equation}\label{IntB}
\int_H {\rm d}{\cal B}=\int_H {\rm d}{\cal B}_c+\int_H {\rm d}{\cal B}_{\rm es}
\end{equation}
and quantum version are meaningful only when the triangulation of $H$ in each term is identical.
Then, for a given triangulation ${\cal T}$,
\begin{eqnarray}\label{IntB_es}
\int_H {\rm d}{\cal B}&=&\sum_{\alpha} \int_{s_\alpha} {\rm d}{\cal B} = \sum_{\alpha} f_\alpha .
\end{eqnarray}
  On the other hand,
\begin{equation}
\int_H {\rm d}{\cal B} \circeq \frac 1 \kappa \int_H \Sigma_{01},
\end{equation}
where $\circeq$ means that the equality is valid on the section $H$ of the IH.
After the same triangulation ${\cal T}$ for $H$, the $(D-2)$-dimensional internal boundary of
a $(D-1)$-dimensional spacelike hypersurface,
\begin{equation}\label{IntSigma01}
\int_H\Sigma_{01}= \sum_{\alpha} \int_{s_\alpha}\Sigma_{01}.
\end{equation}
Therefore,
\begin{equation}\label{f_alpha}
f_\alpha := \int_{s_\alpha} {\rm d}{\cal B} \circeq \frac 1 \kappa \int_{s_\alpha}\Sigma_{01}.
\end{equation}
Eq. (\ref{f_alpha}) provide a boundary condition for bulk spin network
states via the triangulation.
The boundary condition requires that after
  a suitable homotopic transformation,
\begin{enumerate}[1)]
  \item each edge of spin networks in the bulk, which links the bulk and boundary,
  starts or ends at a vertex of a `spin network' for the pure BF theory on
  the horizon;
  \item more than one edge of spin networks in the bulk may approach to a same vertex of a `spin network' for the pure BF theory simultaneously;
  \item an edge happens to lie on a simplex $s_\alpha$ entirely;
  \item an edge does not touch $H$;
 \item a segment of an edge of spin networks in the bulk may
 tangent to a simplex $s_\alpha \in H$, which is divided into two cases:
\begin{enumerate}
 \item it starts or ends at a vertex of a `spin network' for the pure BF theory on
  the horizon and will become 1) after a homotopic transformation;
 \item its pair of vertices are still in the bulk and it will become 4) after a homotopic transformation.
\end{enumerate}
\end{enumerate}
The boundary condition defines the local degree of freedom for the vertices
of `spin networks' for the boundary BF theory on the boundary.  Note that $\int_{s_\alpha} \Sigma_{01}$ gives the flux `area', which must be greater than zero.  Its quantum version
determines the bulk spin network states which is responsible for the boundary degrees of freedom. So next task is to calculate the spectrum of this operator.

Before ending the section, it should be remarked that the information of topology of an IH is
hidden in the triangulation of $H$.  For different topologies, the minimal number of $(D-2)$-simplices will be different.

\section{Quantum states in the bulk and the entropy calculation}
\subsection{Quantum states in the bulk in 4-dimension}
In 4-dimension, the bulk gravitational field can be quantized by the
honolomy of SU$(2)$ connection and the flux of the momentum conjugate to the connection
\cite{al2,hmh1}.
The spectrum of the flux operator for the surface is given by
\cite{{bianchi}}
\begin{equation}\label{flux-on-edge1}
    \int_{S}\hat \Sigma_{01}D^{(j)}(h_{e_\alpha}[{\sf A}])=\pm \kappa \beta
T^{(j)}_{1} D^{(j)}(h_{e_\alpha}[{\sf A}]),
\end{equation}
where $D^{(j)}(h_{e_\alpha}[{\sf A}])$ is the representation $j$ of the holonomy of the SU(2)
connection {\sf A} along the edge $e_\alpha$, which starts or ends on $s_\alpha \in S$,
$T_1^{(j)}$ is a traceless hermitian matrix given by the representation $j$ of the  first generator of SU$(2)$, $\beta$ is the Barbero-Immirzi parameter, and the sign in Eq. (\ref{flux-on-edge1}) is dictated by the relative orientation of the surface $S$ and the edge $e_\alpha$.  In a generic case, the relative orientation is determined by two factors.
One is that the edge starting or ending on $S$ and the other is that the edge lies above or
below $S$.  For the edges which do not touch $S$ or a segment (without a vertex) of
which lies on $S$, $\int_S\hat\Sigma_{01}D^{(j)}(h_e[{\sf A}])=0$.

For a section of an IH, $S=H$, the edges are all on one side of $H$.  Hence, the
sign is only determined by the edge starts or ends on $H$, and the eigenvalue equation satisfied by $\int_{s_\alpha}\hat \Sigma_{01}$ may be written as
\begin{equation}\label{32}
\int_{s_\alpha}\hat \Sigma_{01}|\{m_\gamma\} ,\cdots\rangle
=\kappa \beta m_\alpha |\{m_\gamma\},\cdots \rangle,   \qquad s_\alpha \in \{s_\gamma\},
\end{equation}
where $|\{m_\gamma\},\cdots \rangle$ represent a spin network state in the bulk,
each $m_\gamma \in \{-j_\gamma, -j_\gamma+1, \cdots, j_\gamma -1, j_\gamma\}\ \backslash\ \{0\}$ with $j_\gamma \in \{1/2, 1, 3/2, \cdots\}$.  The zero eigenvalue of $\int_{s_\alpha}\hat \Sigma_{01}$ has to be removed from the spectrum because it gives the flux area of the 2-simplex $s_\alpha$
and because the 2-simplex $s_\alpha$ shrinks to 1-simplex or even a point when the area is zero so that the triangulation changes.  The flux area operator $\widehat{Ar}[H]$ for an IH may be written as \cite{blv1}
\begin{equation}\label{31b}
    \widehat{Ar}[H]=\int_H |\hat \Sigma_{01}|=\sum_\alpha \int_{s_\alpha}|\hat \Sigma_{01}(s_\alpha)|.
\end{equation}
Thus, the flux area $Ar[H]$ for a spin network state is
\begin{equation} \label{Area-eigenvalue}
Ar[H] = \kappa \beta\sum_{\alpha}| m_\alpha|, \quad m_\alpha \in \mathbb{Z}/2\setminus \{0\},
\end{equation}
and is denoted by $a_H$.

\subsection{Quantum states in the bulk in any dimension}

In the Bodendorfer-Thiemann-Thurn (BTT) formulation of quantum gravity in an arbitrary
dimension \cite{{btt1},{btt2},{btt3},{btt4},{btt5}}, even for $D$-dimensional Lorentz
manifold, the internal symmetry SO$(D)$ is still considered.  They introduce $D$ vectors
on a $(D-1)$-dimensional spacelike hypersurface \cite{btt1}.  The $D$ vectors consist of $D$-beins.
To avoid the confusion with the vielbein in Sec. II, we denote them by
$e_{\dot a}^{\hat I}$, where $\dot a=1,\cdots, D-1$ are spatial indices and
$\hat I=1,\cdots, D$ internal indices\footnote{Remember that in the Sec.II,
$a, b, \cdots$ are the abstract spacetime indices, $i, j, \cdots$ are spatial indices
but run from 2 to $D-1$, $I, J, \cdots$ are vielbein indices on $D$-dimensional spacetime.}.
With the help of the $D$-beins, a positive-definite metric on a $(D-1)$-dimensional spacelike
hypersurface,
\begin{equation}\label{spatialmetric}
q_{\dot a\dot b}=\delta_{\hat I\hat J}e_{\dot a}^{\hat I} e_{\dot b}^{\hat J},
\end{equation}
and the $D$th internal vector
\begin{equation}\label{n-vector}
n^{\hat I}:=\frac 1 {(D-1)!}\frac 1 {\sqrt{q}} \varepsilon^{\dot a_1\cdots \dot a_{D-1}}
\varepsilon^{\hat I}{}_{\hat J_1\cdots \hat J_{D-1}}e_{\dot a_1}^{\hat J_1}\cdots e_{\dot a_{D-1}}^{\hat J_{D-1}}
\end{equation}
can be introduced.  Then, $e_{\dot a}^{\hat I}$ and $n^{\hat I}$
span $D$-dimensional internal space with $e_{\dot a}^{\hat I}n_{\hat I}=0$ and $n_{\hat I} n^{\hat I}=1$.

The SO($D$) connection over $(D-1)$-dimensional spacelike hypersurface,
${\mathscr A}_{\dot a \hat I  \hat J}$, is chosen as the configuration variable.
Its canonical momentum is $\pi^{\dot a \hat I \hat J}=2\sqrt{q}q^{\dot a\dot b}
n^{[ \hat I}e_b^{ \hat J]}$.  Define
\begin{eqnarray}\label{E}
  E^{\dot a}_{\hat I} &:=& -\delta_{\hat I \hat J}\pi^{\dot a \hat J \hat K}n_{\hat K} = \delta_{\hat I \hat J}\sqrt{q}q^{\dot a\dot b}e_{\dot b}^{\hat J}, \\
  Q^{\dot a\dot b}&:=&\delta^{\hat I\hat J} E^{\dot a}_{\hat I} E^{\dot b}_{\hat J}, \qquad Q=\det(Q^{\dot a \dot b}).  \label{Q}
\end{eqnarray}
Then, $n^{\hat I}$ and $\pi^{\dot a \hat I \hat J}$ can be expressed in terms of the densitised
vielbein and densitised metric just like in 4 dimension:
\begin{eqnarray}\label{n}
  n^{\hat I} &:=&\frac 1 {(D-1)!}\frac 1 {\sqrt{Q}} \varepsilon_{\dot a_1\cdots \dot a_{D-1}}
\varepsilon^{\hat I \hat J_1\cdots \hat J_{D-1}}E^{\dot a_1}_{\hat J_1}\cdots E^{\dot a_{D-1}}_{ \hat J_{D-1}}\qquad\\
\pi^{\dot a \hat I \hat J} &=& 2 n^{[\hat I}\delta^{\hat J] \hat K} E^{\dot a}_{\hat K}. \label{pi}
\end{eqnarray}
One may further define $Q_{\dot a \dot b}$ and $E^{\hat I}_{\dot a}$ by
\begin{equation}
Q_{\dot a \dot c}Q^{\dot c\dot b}:=\delta^{\dot b}_{\dot a}, \qquad E^{\hat I}_{\dot a} := \delta^{\hat I\hat J}Q_{\dot a \dot b}E^{\dot b}_{\hat J}.
\end{equation}
They satisfy
\begin{equation}
E^{\hat I}_{\dot a} E^{\dot b}_{\hat I}=\delta^{\dot b}_{\dot a}, \qquad
E^{\hat I}_{\dot a} E^{\dot a}_{\hat J}=\delta^{\hat I}_{\hat J}.
\end{equation}
Hence, one may raise and lower spatial indices by $Q^{\dot a\dot b}$ and $Q_{\dot a\dot b}$.
When $D\geq 4$, more constraints, called the simplicity constraints, are needed
in addition to Gauss constraint, spatial diffeomorphism constraint, and Hamiltonian
constraint.

The dual of $\pi^{\dot a \hat I \hat J}$ on a $(D-1)$-dimensional spacelike hypersurface is defined by \cite{btt3}
\begin{equation}\label{pidual}
  (*\pi^{\hat I \hat J})_{\dot a_2\cdots \dot a_{D-1}}=\pi^{\dot a \hat I \hat J}
  \varepsilon_{\dot a \dot a_2\cdots \dot a_{D-1}}.
\end{equation}
The flux of canonical momentum $\pi^{a\hat I\hat J}$ through a $(D-2)$-dimensional spacelike
hypersurface $S$ with a binormal $(n^{[\hat K \hat L]})_{\hat I \hat J}=\delta_{\hat I}^{[\hat K}\delta_{\hat J}^{\hat L]}$ is defined by
\begin{equation}%\label{}
  Fl[S] := \int_S (n^{[\hat K \hat L]})_{\hat I \hat J}\frac {1}{(D-2)!}(*\pi^{\hat I\hat J})_{\dot a_2\cdots \dot a_{D-1}}{\rm d}x^{\dot a_2}\wedge \cdots\wedge {\rm d} x^{\dot a_{D-1}}.
\end{equation}
Then, the flux through $H$ which is coordinated by $\zeta^i$ $(i=2,\cdots, D-1)$ is
\begin{eqnarray}\label{pi-flux}
 \int_H\pi^{D1} &=& \int_H (\delta_{\hat I}^{[D}\delta_{\hat J}^{1]})\frac{1}{(D-2)!}
  (*\pi^{\hat I\hat J})_{i_1\cdots i_{D-2}}{\rm d}\zeta^{i_1}\wedge \cdots \wedge {\rm d}\zeta^{i_{D-2}} \notag \\
  &=& \int_H (n^D\delta^{1\hat K}E^1_{\hat K}-n^1\delta^{D\hat K}E^1_{\hat K}){\rm d}^{D-2}\zeta
  =\int_H\sqrt{\sigma}{\rm d}^{i_{D-2}}\zeta=Ar[H],
\end{eqnarray}
where $\sigma$ is the determinant of the metric on $H$.
It gives the flux `area' for $H$.  Its quantized version should be
\begin{equation}\label{flux-on-edge}
    \int_{H}\hat \pi_{D1}D^{(j)}(h_{e_\alpha}[{\mathscr A}]) =
\pm \frac{1}{2}\kappa \beta R^{(j)}_{D1} D^{(j)}(h_{e_\alpha}[{\mathscr A}])
\end{equation}
where $D^{(j)}(h_{e_\alpha}[{\mathscr A}])$ is now the representation $j$ of holonomy of SO$(D)$
connection $\mathscr{A}$ along the edge $e_\alpha$, which starts of ends on $s_\alpha \in H$,
$R_{D1}^{(j)} $ is the $D1$ generator of SO$(D)$ in the representation $j$,
$\beta$ is the analog of Barbero-Immirzi parameter.
The difference between 4 dimension and an arbitrary dimension is due to the factor that
in an arbitrary dimension SO$(D)$ group only has integer representation, while in 4 dimension
SU(2) can have half-integer representation.  Certainly in 4 dimension SO$(4)$ group can also
be used, which gives different $\beta$ with the SU$(2)$ group in principle, but they give
the same area spectrum, thus the same physics \cite{btt3}.

Again, for edges which do not touch $H$ or lie on $H$ entirely, $\int_H\hat\pi_{D1}D^{(j)}(h_e[A])=0$.  The eigenvalue equation
satisfied by $\int_{s_\alpha}\hat \pi_{D1}$ may be written as
\begin{equation}\label{32}
\int_{s_\alpha}\hat \pi_{D1}|\{m_\gamma\} ,\cdots\rangle
=\kappa \beta m_\alpha |\{m_\gamma\},\cdots \rangle,   \qquad s_\alpha \in \{s_\gamma\},
\end{equation}
where $|\{m_\gamma\},\cdots \rangle$ represent a spin network state in the bulk,
each $m_\gamma \in \{-j_\gamma/2, -j_\gamma/2+1/2, \cdots, -1/2,0,1/2,\cdots
j_\gamma/2-1/2, j_\gamma/2\}\ \backslash \ \{0\}$ with $j_\gamma \in \{1,2,3, \cdots\}$ for SO($D$), $m_\alpha
\in \{m_\gamma\}$ is the value of the $\alpha$ edge which starts or ends at $s_\alpha $ and thus  $m_\alpha \in \mathbb{Z}/2$, and $\cdots$
represents other quantum numbers (say, intertwiners, etc.) that character the bulk state.
The reason that the quantum number $m_\gamma$ jumps by a half is that the half factor in Eq.
(\ref{32}) has been absorbed in the `magnetic' quantum number.  Again,
the zero eigenvalue has to be removed from the spectrum as above.
The flux `area' operator $\widehat{Ar}[H]$ for an IH may be written as %\cite{blv1}
\begin{equation}\label{31c}
    \widehat {Ar}[H]=\int_H |\hat \pi_{D1}|=\sum_\alpha \int_{s_\alpha}|\hat \pi_{D1}(s_\alpha)|,
\end{equation}
and has the eigenvalue for a spin network state, which has the same form as Eq.(\ref{Area-eigenvalue}).

\subsection{Entropy calculation}

In loop quantum gravity approach, the entropy of the IHs is given by
\begin{equation}\label{29b}
    S_{\rm IH}=\ln \mathcal{N}_{\rm IH},
\end{equation}
where $\mathcal{N}_{\rm IH}$ is the number of physical boundary states compatible with constraint 
\begin{equation}\label{global-constraint}
\frac 1 \kappa a_H = \sum_\alpha |f_\alpha| =\beta\sum_{\alpha}| m_\alpha|,\quad m_\alpha \in \mathbb{Z}/2\setminus \{0\}.
\end{equation}

When the constraint on triangulation for different topology is ignored (it will give the sub-sub-leading term), the number of degree of freedom of microstates from (\ref{global-constraint}) is given by
\begin{equation}\label{36}
     \mathcal{N}_{\rm IH} =\sum_{n=1}^{n=2 a} C_{2 a-1}^{n-1} 2^n=2\times 3^{2 a-1} ,
\end{equation}
where $a:=a_H/(\kappa \beta)$.
The entropy is given by
\begin{equation}\label{38}
    S_{\rm IH}=\ln\mathcal{N}_{\rm IH}=2 a \ln3+\ln \frac{2}{3}= \frac{ \ln3}{\pi \beta}\frac{a_H}{4G}+\ln \frac{2}{3}.
\end{equation}
Its leading term just gives the area law. When $\beta$ is set to $\beta=\frac{\ln 3}{\pi}$, the famous coefficient $\frac{1}{4}$ can be achieved.

\section{Relax of gauge fixing condition}

The choices of  orthogonal co-vielbein (\ref{6}) near the IH in $D$-dimensional spaceetime fix the
gauge SO(1,1)$\times$SO($D-2$) from Lorentz group SO($D-1$,1).  Hence, the conclusions about the symplectic form and entropy seem to depend on the gauge fixing.  On the other hand, it is shown in Ref. \cite{bcg1} that an IH  will reduce the local Lorentz group to its subgroup $ISO(D-2)\ltimes SO(1,1)$. It is this SO$(1,1)$ group that is used
for the BF theory. The following calculation shows that the conclusion is independent of the gauge chosen for this subgroup.

Near an IH in $D$-dimensional spacetime, one may always set up a vielbein $(l^a, n^a, e_{\tt A}^a)$ as described in the Sec.II and the inverse metric of the spacetime is
\begin{equation}\label{metric}
  g^{ab}= -l^a\otimes n^b - n^a\otimes l^b +\delta^{\tt AB}e_{\tt A}^a\otimes e_{\tt B}^b.
\end{equation}
The inverse metric (\ref{metric})
takes the same form under the following local transformations (cf. \cite{bcg1}),
\begin{equation}
%\left \{
\begin{array}{lll}
l^a \mapsto \frac 1 \alpha l^a,       & n^a \mapsto \alpha n^a,  & e_{\tt A}^a \mapsto e_{\tt A}^a, \vspace{0.6em} \\
           l^a \mapsto  l^a, & n^a \mapsto n^a, & e_{\tt A}^a \mapsto \Lambda^{\tt B}{}_{\tt A}e_{\tt B}^a,
              \vspace{0.6em} \\
           l^a \mapsto  l^a,  & n^a \mapsto n^a - b^{\tt A} e_{\tt A}^a +\frac 1 2 b^2 l^a, &
           e_{\tt A}^a \mapsto e_{\tt A}^a -  b_{\tt A} l^a , \vspace{0.6em}\\
           l^a \mapsto l^a - w^{\tt A} e_{\tt A}^a +\frac 1 2 w^2 n^a, & n^a \mapsto  n^a,  &
           e_{\tt A}^a \mapsto e_{\tt A}^a -  w_{\tt A} n^a ,
         \end{array} %\right .
\end{equation}
where $\Lambda^{\tt B}{}_{\tt A} \in $ SO$(D-2)$, $b^2=\delta_{\tt AB}b^{\tt A}b^{\tt B}$, $b_{\tt A}=b^{\tt A}$,
$w^2=\delta_{\tt AB}w^{\tt A}w^{\tt B}$ and $w_{\tt A}=w^{\tt A}$.  These transformations form
an SO$(D-1,1)$ group.  It is shown that only the first three types of transformation can keep the isolated horizons boundary condition invariant \cite{bcg1}. So only those transformations are considered.

The first two classes of transformations belong to SO(1,1)$\times$SO($D-2$).  Undoubtedly, the calculation in Sec.II. is invariant under the two classes of transformations.

For the third class of transformations, a set of orthogonal vielbein fields which are compatible with the metric may be chosen as
\begin{eqnarray}
 &&   E_0^a=-\sqrt{\frac{1}{2}}[\alpha  (n^a - b^{\tt A} e_{\tt A}^a +\frac 1 2 b^2 l^a)  +\frac{1}{\alpha}  l^a ], \nonumber \\
 &&\label{52}    E_1^a =\sqrt{\frac{1}{2}}[\alpha  (n^a - b^{\tt A} e_{\tt A}^a +\frac 1 2 b^2 l^a)  -\frac{1}{\alpha} l^a  ],\\
 &&   E_{\tt A}^a=e_{\tt A}^a -b_{\tt A} l^a, \nonumber
\end{eqnarray}
where $\alpha(x)$ is an arbitrary function of the coordinates. The covielbein fields
dual to the above vielbein are given by
\begin{eqnarray}\label{50}
  &&  E^0_a =\sqrt{\frac{1}{2}}[\alpha  (n_a - b_{\tt A} e^{\tt A}_a +\frac 1 2 b^2 l_a) +\frac{1}{\alpha}  l_a ],\nonumber \\
  &&  E^1_a=\sqrt{\frac{1}{2}}[\alpha  (n_a - b_{\tt A} e^{\tt A}_a +\frac 1 2 b^2 l_a) -\frac{1}{\alpha} l_a]), \\
  &&      E^{\tt A}_a=e^{\tt A}_a -b^{\tt A} l_a= e^{\tt A}_{\mu}dx^{\mu}-b^{\tt A} l_\mu dx^\mu. \nonumber
\end{eqnarray}
Obviously,
$(E_0,E_1)$ with different choices $\alpha(x)$ are also related by a $1+1$ Lorentz transformation, and  on the horizon
\begin{equation}\label{51}
    E_0\triangleq E_1.
\end{equation}
With the same calculation, it can be shown that the connection also satisfy the higher dimensional correspondence of (\ref{6}), and thus the same conclusion will be drawn.

Therefore, the final results are independent of the gauge chosen for the residual gauge symmetry of IHs.

%%%%%%%%%%%%%%%%%%%%%%%%%%%%%%%%%%%%%%%%%%%
%%%%%%%%%%%%%%%%%%%%%%%%%%%%%%%%%%%%%%%%%%%
%%%                                     %%%
%%%            Discussion               %%%
%%%                                     %%%
%%%%%%%%%%%%%%%%%%%%%%%%%%%%%%%%%%%%%%%%%%%
%%%%%%%%%%%%%%%%%%%%%%%%%%%%%%%%%%%%%%%%%%%

\section{Discussion}
In this paper, a coupled system is considered, whose bulk is described by the Palatini action and whose internal boundary (which is an isolated horizon) is governed by an SO$(1,1)$ BF theory. Due to the
existence of the boundary action, the symplectic form contains just the bulk term.
So, after quantization only the bulk Hilbert space is present. The boundary condition chooses the available {\it bulk} spin network states.  Their internal degrees of freedom are traced out and the number of their boundary degrees of freedom gives the entropy of the IH.  In contrast, in the previous approach \cite{wmz,wh1,whl1,wh2} since only the Palatini action is considered, the symplectic form has two terms, the bulk term and the boundary term. The resulted Hilbert space is also the tensor product of the bulk and the boundary Hilbert space. The boundary condition picks out the available {\it boundary} states. The bulk states are traced out and the number of the independent available boundary states gives the entropy of the IH. In this sense, the reformulated approach is somewhat similar to Rovelli's original idea, but one of the essential differences from the Rovelli's idea is that a boundary BF action is added to cancel the boundary symplectic current to get a conserved symplectic form on the Cauchy hypersurface $M$.

On an IH, the boundary BF theory coupled with a Palatini gravity may be classically decomposed into a pure BF
theory and a fixed interaction term determined by the Palatini gravity.  The pure BF theory,
as a free BF theory, can be quantized first through loop quantization \cite{bf1}.   The set of the `generalized spin network states'
in the loop quantization, as the basis of the Hilbert space over the IH, define
the set of triangulations for the section of the IH.  The boundary BF fields are expanded according to the set of triangulations and set up the connection with the bulk spin network states.  Since the definition of the bulk spin network states are completely presented by loop quantization of gravity in the bulk  and are independent of the quantized boundary BF theory, the role of the quantized boundary BF theory in the reformulated approach is just to provide a set of triangulations to select the bulk spin network states.
As a by-product, the different topologies of isolated horizons appear in the sub-sub-leading term of the entropy expression.

Although the two boundary BF theory approaches have different structures, they have some  common features.
Both of them can be applied to IHs in any dimensional spacetime.  This is
the distinctive advantage over other approaches in the framework of loop
quantization, which are either limited in 4 dimensional spacetimes \cite{{ans1},{tt1},{rov2}}
or limited in even dimensional spacetimes \cite{{abck},{rot1},{rot2},{btt5},{bodendorfer}}.
The SO$(1,1)$ group exist for all isolated horizons,
no matter what kind of other symmetry they have, and
even no matter whether or not they have other symmetry.  Therefore, the method can be further generalized easily to deal with the
entropy of black branes or other black objects.
 On the other hand, it is vital important in the approaches that IHs serve as the internal boundary of the spacetime because only an IH does permit an SO(1,1) boundary BF theory.
It means that by use of the SO$(1,1)$ boundary BF theory one can only calculate the entropy
of a section of an IH.  It seems a limitation of the approaches.  But, due to
the limitation, one does not needed to explain why an arbitrary surface (or hypersurface) has the area entropy just as an horizon and what the physical meaning of the entropy is for an arbitrary surface or hypersurface, which have to be faced in the approaches \cite{{ans1},{tt1},{rov2},{abck},{rot1},{rot2},{btt5},{bodendorfer}}.
In addition, in the boundary BF theory approaches, the flux-area operator and thus the simplified area spectrum
are used, instead of area operator itself and the full spectrum of the area operator.
It is also an advantage because in a loop quantization of a generalized gravity the
flux-area operator turns out to measure the Wald entropy \cite{bn1,whl1}.

It is interesting to compare the boundary BF theory approaches with Carlip's conformal field
theory explanation \cite{carlip1,carlip2}.  In the boundary BF theory approach, the full
geometry is considered, while in the conformal field approach the 2-dimensional geometry
near horizon is dealt with.  In the former, the internal SO(1,1) symmetry is relevant, while
the spacetime conformal symmetry is the key one in the latter.  In the former, a boundary
BF theory is required on the internal boundary, while a conformal field theory, which is
not yet very clear except in 3-dimensional spacetime \cite{carlip3}, is needed on the
internal boundary.  The former has an undetermined coefficient $\beta$ as other loop quantum
gravity approach, while the latter can give the exact Bekenstein-Hawking entropy.  The former
can give the Wald entropy except Bekenstein-Hawking entropy, while the Wald entropy has not
yet been discussed in the latter approach, to our knowledge.  It is worth to explore whether there is an internal connection
between the two approach.

Finally, it should be remarked that the value of $\beta$ obtained here is the same as
the value in Ref. \cite{wmz} but different from the value in Ref. \cite{wh1} with a factor 2.
The reason is that the eigenvalue equation (\ref{flux-on-edge1}) for the flux operator is different from the Ref. \cite{wh1} with a factor 2, so the flux operator have the same eigenvalues, thus the same physics.  We choose this form since it gives the same value for both SU$(2)$ in 4 dimension and SO$(D)$ in arbitrary dimension.
%%%%%%%%%%%%%%%%%%%%%%%%%%%%%%%%%%%%%%%%%%%
%%%%%%%%%%%%%%%%%%%%%%%%%%%%%%%%%%%%%%%%%%%
%%%                                     %%%
%%%            Appendix                 %%%
%%%                                     %%%
%%%%%%%%%%%%%%%%%%%%%%%%%%%%%%%%%%%%%%%%%%%
%%%%%%%%%%%%%%%%%%%%%%%%%%%%%%%%%%%%%%%%%%%

\section*{Acknowledgments}
We would like to thank Prof. Y.-G. Ma for critical and helpful discussion.
This work is supported by National Natural Science Foundation of China under the grant
11275207.% and by China Postdoctoral Science Foundation.
%\section*{References}
\bibliography{rotating150520}
\end{document}